# Optical Modulation Due to Energy Exchange Between Photonic and Exciton Modes in the Intermediate Coupling Regime

*Evripidis Michail [1, 2, 3], Sander A. Mann [1, 2, 4], Kamyar Rashidi [1, 2], Christoph Lambert [5], Vinod M. Menon [1, 6], Andrea Alù [1, 2], Matthew Y. Sfeir [1, 2, 3, *]*

[1] Department of Physics, Graduate Center, City University of New York, New York, NY 10016, USA,

[2] Photonics Initiative, Advanced Science Research Center, City University of New York, New York, NY 10031, USA,

[3] School of Chemistry and Biochemistry, School of Materials Science and Engineering, Georgia Institute of Technology, Atlanta, Georgia 30332, USA,

[4] Institute of Physics, University of Amsterdam, Amsterdam, The Netherlands,

[5] Institut für Organische Chemie, Universität Würzburg, Würzburg D-97074, Germany,

[6] Department of Physics, Center for Discovery and Innovation, City University of New York, New York, NY 10031, USA

[*] Corresponding Author: msfeir@gatech.edu

**ABSTRACT:** Actively tunable photonic devices are vital for next-generation optoelectronics requiring rapid switching and high bandwidth. Although organic optoelectronic devices have found wide application, their use as optical modulators has been limited by low absorption in the critical near-infrared (NIR) region, slow response time, and weak nonlinearities. To address these limitations, we developed a scheme based on intermediate exciton-photon coupling in a NIR absorbing squaraine-dye based photonic structure. Using energy-momentum resolved pump-probe spectroscopy, we show that the sign and magnitude of the optical response of our system depends strongly on the energy detuning between the excitonic and photonic modes. These data are analyzed using temporal coupled-mode theory to show that near resonance, a distinct energy exchange process emerges in the cross-over regime between strong and weak light-matter coupling. This effect enables dynamical control over the photoinduced response, providing a pathway for broadband optical signal modulation extending into the NIR spectral region.

## INTRODUCTION

Organic optoelectronics is an emerging field of significant research growth and technological interest spanning light emitting diodes, photovoltaics, and light detectors.[1, 2, 3] Compared to inorganic platforms, organics offer high mechanical flexibility, simple processing, cost-effective recyclability and easy tunability.[4, 5, 6] A key challenge is the development of devices whose optical response can be modulated in a reversible and real time manner for all-optical technologies, such as quantum information processing, ultrafast optical switching, sensing and imaging.[7, 8, 9, 10, 11] Furthermore, enabling the operation of these optical systems in the NIR spectral regime would minimize optical absorption and scattering losses, permitting high energy efficiency and resolution across a wide range of technological applications.[12, 13] Realizing this goal requires both the development on new organic materials with strong NIR absorption and device architectures that permit rapid and efficient signal modulation in the presence of a low-power optical field and without strong external stimuli (e.g., magnetic or RF fields).[14, 15]

Several recent breakthroughs have demonstrated that organic chromophores can be developed with strong NIR absorption and enhanced optical nonlinearities and long response times. One approach to the development of organic optical modulators and molecular switches has focused on conducting polymers, such as PEDOT:PSS (poly(3,4-ethylenedioxythiophene) polystyrene sulfonate), EDOT (3,4-ethylenedioxythiophen), $PProDOTMe_2$, Polyaniline, bisthienylbenzen, which have been employed as active media in plasmonic nanostructures.[16, 17, 18, 19] However, as it is challenging to dynamically alter the material composition or geometry of the nanostructure in a reversible and real-time manner, these approaches typically rely on strong external stimuli such as high intensity light, heat, or electric potential to modulate the optical properties of the device.[20] These trigger responses often require a complex fabrication process and induce non-reversible changes in the material due to cycling that can degrade the functionality of the structure over time.[21, 22, 23] Alternatively, small molecule NIR organic dyes have recently been developed that exhibit strong oscillator strengths, large exciton delocalization, and large intrinsic nonlinear optical coefficients. However, small molecules have not been broadly examined in the context of optical switches as schemes to dynamically modulate their optical properties with high efficiency have not been reported.

An innovative and widely proposed strategy to amplify and macroscopically tune the optical responses of molecular systems is to strongly couple organic molecules to a confined electromagnetic field.[24] In this regime, coherent energy exchange occurs between the excitonic and photonic components with a rate that exceeds energy dissipation from the system. This gives rise to hybridized exciton-photon states (**Figure 1a**) with a characteristic Rabi splitting energy, $\hbar\Omega_R$.[25] Recent work has demonstrated promising schemes for the efficient manipulation of the excited state properties of organic systems, including numerous phenomena in optoelectronics.[26, 27] For example, the ability to tune the relative weights of the photonic and excitonic states through control of the light-matter coupling strength ($g = 2\hbar\Omega_R$) is a key capability for optical switches.[28, 29] This includes the development of electrochemical devices that can switch between the weak (**Figure 1b**) and strong coupling regime using an external bias.[30, 31, 32, 33] However, this approach is relatively slow, with switching times on the order of seconds.[34] In inorganic semiconductors, effective ultrafast nonlinear switching between strong and weak coupling between an intersubband transition and a cavity mode has been demonstrated using ultrafast optical pulses.[35] However, this scheme is more challenging to implement in organic materials, as their broader intrinsic linewidths and overlapping transitions results in smaller nonlinearities.[36] These properties also make the crossover between the weak and strong coupling regimes difficult to experimentally identify. As a result, the fundamental physics describing the transition between weak to strong coupling remains insufficiently studied in organics, and schemes to exploit this phenomenon have not been reported to date.

Here, we demonstrate that we can exploit the dynamical properties of organic photonic devices to enhance its optical nonlinearities in the intermediate coupling regime, where the light-matter interactions are distinct from either the strong and weak coupling regimes (**Figure 1c**). We demonstrate a system for which intermediate coupling is achieved in the NIR with an "open cavity" metal – organic dielectric layer photonic system (PS) and a NIR absorbing squaraine small-molecule chromophore (**Figure 1d**). By employing a high-sensitivity momentum-resolved transient optical spectrometer, we map the optical response of the PS for different in-plane momenta of the incident probe beam. Our data demonstrate that the excited state character depends strongly on the detuning, $\delta = \hbar\omega_{ph} - \hbar\omega_{ex}$, between the resonant energy of the waveguide (WG) mode $\hbar\omega_{ph}$ and the exciton resonance energy $\hbar\omega_{ex}$. For positive ($\delta > 0$) and negative ($\delta < 0$) detuning, the light and matter components appear to respond independently to optical pumping. In contrast, near the degeneracy ($\delta \approx 0$), the response of the system to optical pumping indicates two strongly coupled species that interfere with each other, leading to an electromagnetically induced transparency effect[37, 38, 39] Using temporal coupled-mode theory (TCMT) model,[40, 41] we show that our results can be understood in the context of the dynamical interaction between three fundamental resonances – two broad excitonic transitions and a photonic WG mode – that exchange energy near the degeneracy point. Although these resonances remain distinct and can be directly excited with external optical pumping, they induce a transient response in the other resonances through their internal coupling. This can be contrasted with strongly coupled systems, in which hybridized states dominate the optical response. We establish that dynamical control of the sign of the photoinduced response of the system is a hallmark of the intermediate coupling regime.

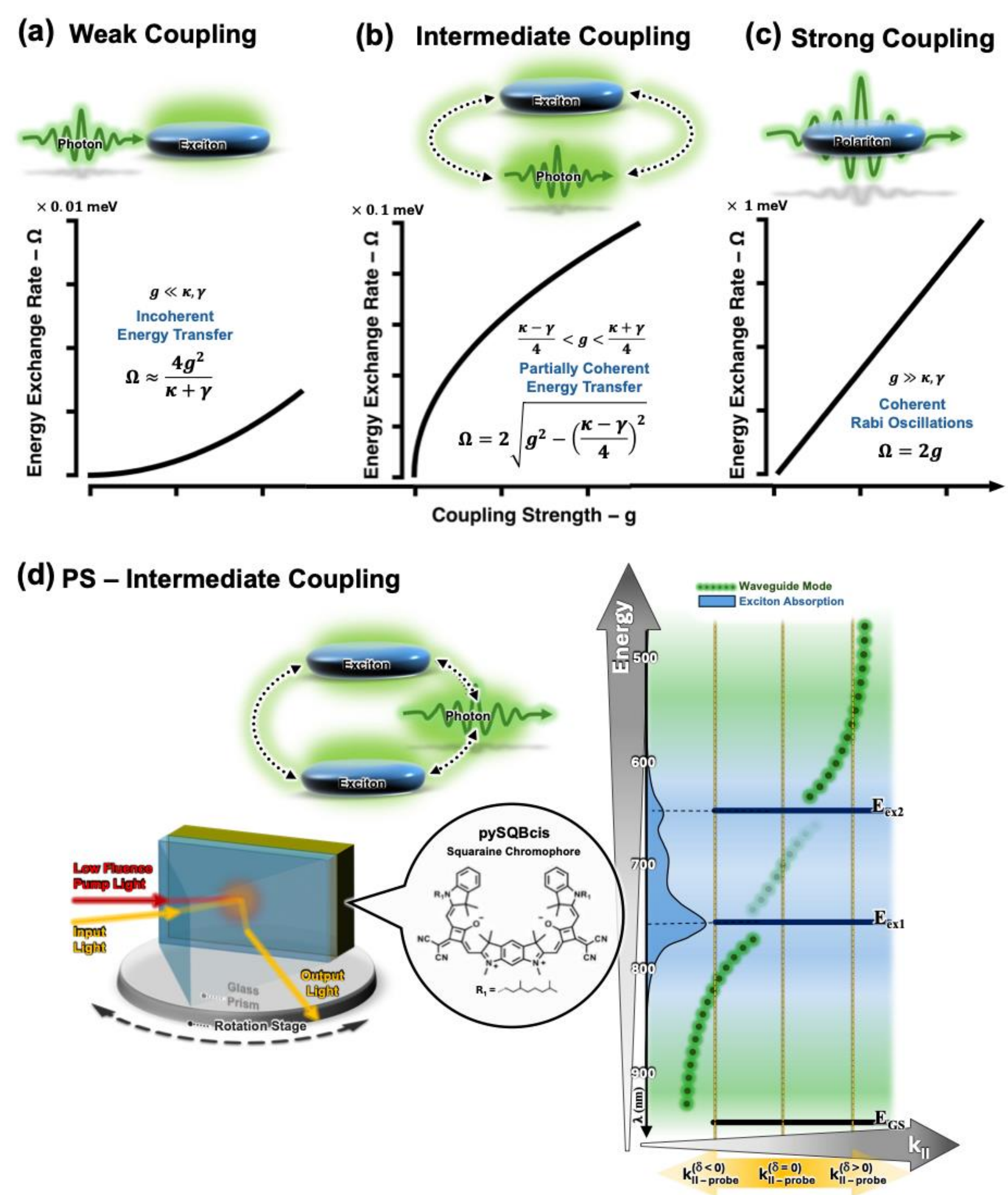

**Figure 1.** Illustration of light-matter coupling regimes, showing the energy exchange rate Ω as a function of the coupling strength g, with γ and κ denoting the exciton and cavity (photon) decay rates, respectively: **(a)** Weak coupling, where distinct matter energy states interact with the incident electromagnetic field. **(b)** Intermediate coupling, where matter and photon energy states

exchange energy when they become nearly degenerate. **(c)** Strong coupling, in which the system consists of well-defined energy states with hybrid light and matter character. **(d)** We employe a PS, based on the Squaraine organic chromophore pySQBcis, consisting of an open-cavity metal organic dielectric layer. Using the Kretschmann-Reather attenuated total reflection configuration, we probe the transient response of the system across a wide range of in-plane momenta $k_{||\text{-probe}}$ under low-fluence photoexcitation. A representative energy-momentum map of the PS which enables intermediate exciton-photon coupling and demonstrates a key dispersion relation, together with the energy distribution of the molecular excitonic states ($E_{ex1}$ and $E_{ex2}$). The yellow vertical lines highlight the intersections of the probe pulses, which correspond to in-plane momenta, $k_{||\text{-probe}}(\theta_{probe})$ where we probe the dynamics of the PS under red ($\delta > 0$), resonant ($\delta \approx 0$) and blue ($\delta < 0$), energy detuning ($\delta = E_{WG} - E_{ex}$) relative to the energies of the excitonic manifold $E_{ex}$.

## RESULTS AND DISCUSSION

Here, we examine squaraine molecules as promising candidates for exploring energy exchange in the intermediate coupling regime, for applications in NIR modulation. Compounds based on squaraine are particularly promising for NIR optoelectronic applications,[42] as they feature simple synthetic access, high photostability, and their broadly tunable optical properties.[43, 44] Here, we study the squaraine dimer pySQBcis (**Figure 1d**),[45] which exhibits a broad exciton linewidth and strong vibronic progression in the NIR spectral region that is typical of this class of materials in both isolated molecules and molecular aggregates.[46, 47, 48,49]. The absorption spectrum of pySQBcis is dominated by two exciton electronic transitions, an $S_0 \rightarrow S_1$ transition near 750 nm and a higher energy exciton absorption near 650 nm has been previously assigned to the $S_0 \rightarrow S_2$ transition (**Figure 2a**). This material features a broad NIR photoluminescence spectrum that extends past 900 nm (**Figure SI-5**). In general, the broad spectral linewidths and multiple overlapping exciton transitions of squaraines have either entirely suppressed the emergence of strong light-matter coupling or significantly modified the nonlinear response of systems close to the strong-coupling limit.[50, 51] However, its strong oscillator strengths are suitable for studies focused on the intermediate coupling regime, which requires that the Rabi splitting satisfies the following condition relative to the linewidths of the exciton transition, $\Gamma_{ex}$, and the photon mode, $\Gamma_{ph}$:[52, 53, 54]

$$\left(\Gamma_{ex} - \Gamma_{ph}\right)^2 < (\hbar\Omega_R)^2 < 2\left({\Gamma_{ex}}^2 + {\Gamma_{ph}}^2\right) \quad \text{(Eq. 1)}$$

The conventional approach employs dyes for which the Rabi splitting can be readily tuned into the strong coupling regime in the limit of high dye concentration. However, pySQBcis is characteristic of a large class of organic chromophores for which strong coupling is difficult to achieve due to their broad absorption linewidths ($\Gamma_{ex}$), despite their large extinction coefficients. This gives a large parameter space and strong optical response to manipulate and study the dynamics of energy exchange between excitons and photons.

We have designed our PS based on an open cavity platform that exhibits beneficial dispersion characteristics and a high quality factor. We have recently shown how these designs can be exploited to allow for energy- and momentum-selective excitation and dynamical characterization of exciton-polaritons.[55] Here, we have designed and fabricated a thin film metal-dielectric stack with pySQBcis as the active organic material. The photonic structure is fabricated by depositing an organic layer consisting of pySQBcis in a PMMA matrix on a BK7 substrate coated with 2 nm of germanium and 40 nm of silver. As previously shown, this PS with thicker organic layers (~ 250 nm) supports waveguide modes (WG) (SI **Figure SI-6** and **SI-7**) polarized opposite to conventional surface plasmon polaritons (TM vs. TE), and feature a sharper dispersion and higher Q factors.[55] In previous studies, strong coupling has been achieved using WG modes and visible-absorbing organic dyes with strong and narrow exciton resonances.[52, 56]

In the pySQBcis-based PS, we observe modifications of the cavity mode dispersion, but do not unambiguously achieve strong light-matter coupling, even at high chromophore concentrations. The effect of pySQBcis incorporation on the dispersion of the cavity mode as a function of concentration is calculated using the transfer matrix method (TMM). Above the light cone, we observe a WG mode whose frequency increases monotonically across the visible and NIR with increasing in-plane momentum. At high concentration, a break in the dispersion is observed near the exciton resonance transitions ex1: $S_0 \rightarrow S_1$ and ex2: $S_0 \rightarrow S_2$ (**Figure 2a**). However, clear signatures of strong coupling are not found, which would be characterized by a well-defined asymptotic behavior of the polariton mode near the exciton resonances and a sharp, well defined middle polariton branch in between. Instead, we observe a broad, indistinct region between the two exciton resonances, whose width increases with increased concentration (**Figure SI-9)**. Outside this region, the photonic mode is relatively unaffected. Experimental angle-resolved reflectivity measurements from sample with a doping concentration of 17 mM in a 250 nm thick polymer matrix (**Figure 2c**) confirm the predicted behavior and show the dispersion of the photonic mode interrupted by a region with poorly defined reflectivity peaks spanning the excitonic resonances (~ 650 – 750 nm). The broad ensemble exciton linewidth ($\Gamma_{ex}$ = 460 meV) of this device is comparable to the calculated Rabi splitting energy from TMM ($\hbar\Omega_R$ = 500 meV). Together, the indistinct dispersion and the similar characteristic energy scales (Eq.1) indicate the intermediate coupling regime, consistent with previously established protocols.[57, 58, 59]

The essential characteristics of our experimental reflectivity spectra are reproduced using a time-domain model based on TCMT, which has previously been shown to accurately describe the dynamics that lead to interference phenomena in coupled oscillators.[41, 60] Importantly, while frequency-domain TMM approaches satisfactorily reproduce steady-state reflectivity measurements, they cannot be used to describe the essential interaction dynamics that are necessary for interpreting key features in the steady-state and transient optical spectra in this regime.[51] As such, we have parameterized the TCMT model to describe the interaction of two excitonic modes, with resonance frequencies $\Omega_1$ ($S_0 \rightarrow S_1$) and $\Omega_2$ ($S_0 \rightarrow S_2$), and a dispersive photonic WG mode $\Omega_3(k_{\|})$. The time-dependent interaction of these resonances with each other and with the incident light field is described as:

$$\frac{d}{dt}\begin{pmatrix} \mathrm{a} \\ \mathrm{m}_1 \\ \mathrm{m}_2 \end{pmatrix} = [i\boldsymbol{\Omega} - \boldsymbol{\Gamma}_r - \boldsymbol{\Gamma}_{\mathbf{a}}]\begin{pmatrix} \mathrm{a} \\ \mathrm{m}_1 \\ \mathrm{m}_2 \end{pmatrix} + \mathbf{d}^{\mathrm{T}} s^{(+)} \qquad \text{(Eq. 2)}$$

Here, the complex amplitudes of the WG mode (a) and the two molecular transitions ($m_1$, $m_2$) are normalized such that their absolute value squared describes the stored energy. The driving amplitude $s^{(+)}$ is normalized such that the absolute value squared gives the input power. The matrices $\boldsymbol{\Gamma}_{\mathbf{a}}$ and $\boldsymbol{\Gamma}_r$ describe the absorption and radiative decay of the system, and the matrix $\mathbf{d}$ describes the coupling between the resonances and incident light field. The analytical forms of the matrices in Eq. 2 are provided in the SI. Assuming nonzero losses for the WG mode due to scattering losses, the reflected amplitude $s^{(-)}$ is given by the expression

$$s^{(-)} = -s^{(+)} + \mathbf{d}\begin{pmatrix} a \\ m_1 \\ m_2 \end{pmatrix} \qquad \text{(Eq. 3)}$$

This simple model reproduces the essential features of our steady-state reflectivity spectra (**Figure 2d**), showing an indistinct broadened region between the exciton resonances. This suggests that we can extend this model to interpret transient optical measurements if we account for changes in the amplitudes of the ensemble of the exciton resonances ($\Omega_1$, $\Omega_2$) due to bleaching of the molecular transitions.

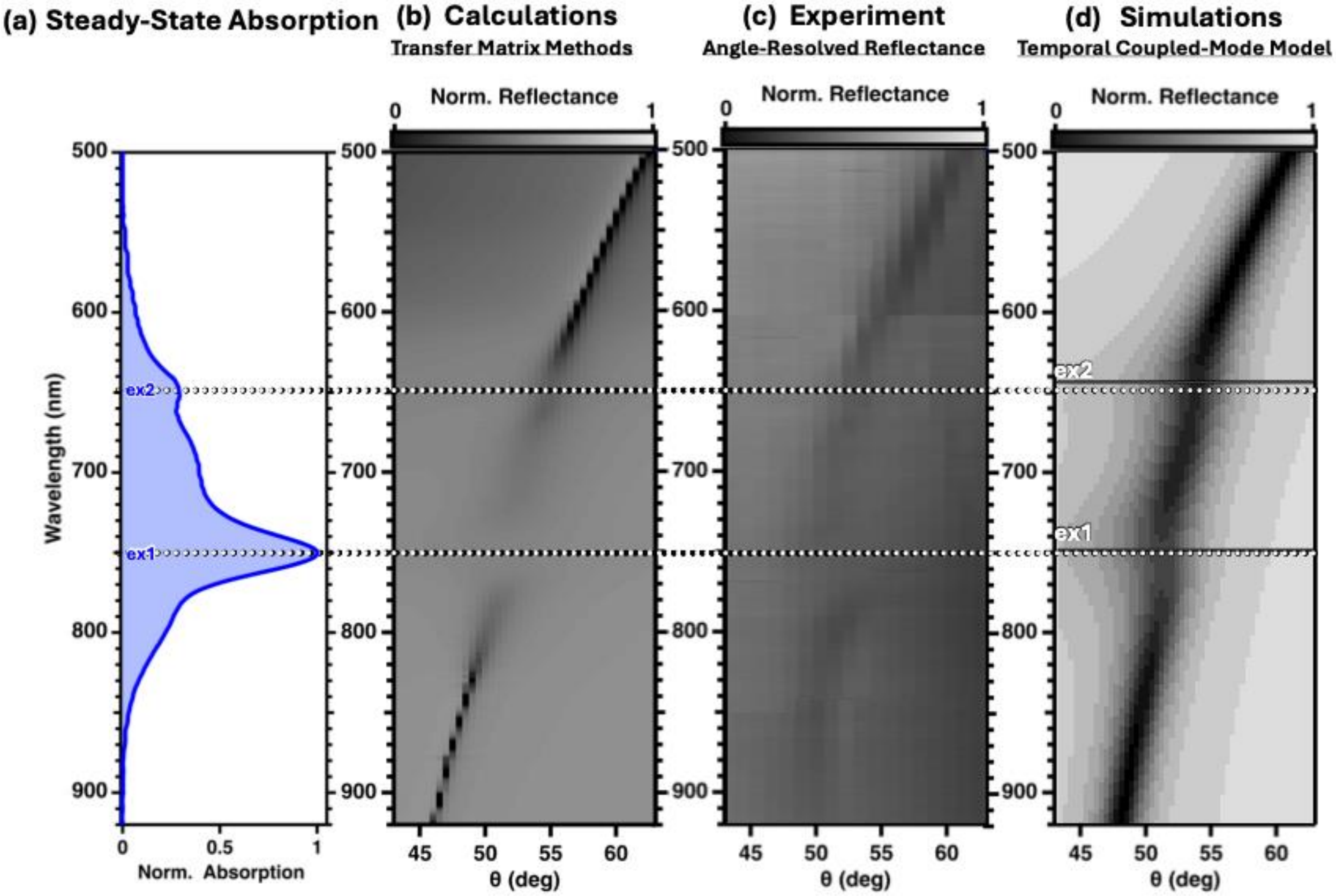

**Figure 2: (a)** Steady-state absorption spectrum of the bare organic dielectric layer indicating the two exciton transitions ex1: $S_0 \rightarrow S_1$ and ex2: $S_0 \rightarrow S_2$ (white dotted lines). **(b)** Calculated angle-reflectivity spectra map obtained by modeling the PS using TMM calcualtions. **(c)** Experimental angle-resolved reflectivity data obtained using Fourier space imaging method. **(d)** Reflectance as function of the incident angles as simulated from TCMT.

Transient optical reflection spectroscopy on the PS was carried out using our high-sensitivity energy - momentum resolved pump-probe spectrometer, based on the Kretschmann-Raether configuration, with low excitation fluence ($F_{pump} \approx 10$ uJ/cm$^2$). This approach allows us to resonantly excite distinct modes of the system by tuning the energy and angle of incidence of the pump beam, including the first ($S_1$ with $\lambda_{pump} = 750$ nm) and second ($S_2$ with $\lambda_{pump} = 650$ nm) excitonic transition of the pySQBcis chromophore, as well as the WG mode at energies far from the excitonic resonances. The pump and probe beams are coupled to the PS through a glass prism with an angular offset $\Delta = \theta_{pump} - \theta_{probe}$. We vary the angle of the prism, with Δ fixed at 4°, to map the transient optical response (−ΔR/R) of the PS at selected wavevectors. For each pump condition, we probe the dynamics of the system for red ($\delta < 0$), resonance ($\delta \sim 0$) and ($\delta > 0$) energy detuning from the excitonic resonances (**Figure 3**). Spectroscopic identification of the WG mode resonance during the transient measurements is achieved by tracking sharp dips in the steady-state reflected probe spectrum as a function of $\theta_{probe}$ (see Figure **SI-12a**).[55, 61]

The temporal evolution of the transient reflectivity, $-\Delta R/R = [R'(\lambda, t) - R(\lambda)] / R(\lambda)$, is used to quantify changes in the refractive index of the sample, where R′ denotes the time-dependent reflectance following photoexcitation. By identifying the various contributions to the transient signal and their individual dynamics, we can interpret the energy exchange between the photonic and excitonic mode in the PS.[62] The individual contributions from the exitonic and photonic components are well understood if they are assumed to be independent. In fact, this is exactly what is observed when we resonantly pump the $S_0 \rightarrow S_2$ transition at 650 nm and probe the dynamics of the PS at in-plane momenta corresponding to negative $\delta < 0$ (**Figure 3a**) and positive $\delta > 0$ (**Figure 3e**) energy detuning from the exciton resonance. For example, contributions from the organic layer are identified by the non-dispersive and broad transient signal that appears within the energy range of the excitonic states ($\lambda_{probe} \sim 650 - 750$ nm). As the organic layer is inherently weakly reflective, changes in its net absorption modulate the near-unity reflection from the silver substrate. If the organic layer becomes less absorptive near the exciton resonance upon photoexcitation (ground state bleaching), we will observe higher net reflectivity ($-\Delta R/R > 0$) from the PS. This response is similar to the transient spectral linewidth of the bare dielectric layer measured under the same excitonic excitation.

The transient response of the photonic modes also appears to be independent from the excitons at probe wavevectors that correspond to wavelengths far from resonance, for example at 890 nm for $\delta < 0$ ($\theta_{probe} \sim 47°$) (**Figure 3a**) and 580 nm for $\delta > 0$ ($\theta_{probe} \sim 58°$) (**Figure 3e**). The spectral dynamics of the photonic modes are characterized by a momentum-dispersive transient signal exhibiting a derivative-shaped $\Delta R/R$ feature that results from a change on the WG mode resonance due to photoexcitation of the PS. These two antisymmetric lobes result from the transient blueshift of the branch of the dispersion located below the exciton resonance energy and a redshift of the branch located above the exciton resonance. This effect occurs since photoexcitation of a subset of the exciton population reduces the number of oscillators coupled to the cavity, lowering the Rabi splitting energy ($\hbar\Omega_R \propto N^{1/2}$)[63, 55]. Here, the direction of the shift is indicated by relative spectral position of the lobe for which $-\Delta R/R > 0$. For example, if the positive lobe appears to the low energy side of the resonance (zero-crossing) then photonic mode red-shifts in the excited state.

The behavior and resulting transient signals of these modes at in-plane momenta of the probe pulses in which the excitonic and photonic modes are nearly energetically degenerate ($\delta \sim 0$) are more complex and not readily described using this non-interacting framework. In this regime, the photonic and excitonic transient components exhibit spectral overlap, but remain distinguishable due to the dispersive nature of the WG mode (see **Figure SI-13**). We assign the derivative-shaped transient feature originates from the WG mode centered at $\lambda_{probe} \sim 770$ nm, which retains its transient blueshift spectral character (**Figure 3b**). Surprisingly, the transient signal associated with the organic layer response at energies resonant with the molecular transition exhibits the opposite optical behavior (**Figure 3b**). Indeed, the transient signal at energies resonant with the broad exciton manifold (700 – 750 nm) exhibits a positive sign ($-\Delta R/R > 0$) at near-resonance energy detuning ($\delta \sim 0$), in contrast to its behavior ($-\Delta R/R < 0$) when probed at large energy detunings ($\delta < 0$ and $\delta > 0$). Importantly, this indicates that the PS is becoming more absorbing at energies that would normally be associated with molecular ground state bleaching. This behavior is inconsistent with the conventional picture of molecular photophysics, which suggests that an out-of-equilibrium system should exhibit an induced transparency near the exciton resonance, increasing the overall reflectivity signal from the metal substrate. However, the opposite behavior emerges here due to interactions with the photonic modes of the system.

We explain this anomalous behavior in the context of the intermediate coupling regime, in which optical excitation of either the discrete WG mode and the excitonic states will modify the local refractive index due to rapid energy exchange. Furthermore, a characteristic of this regime is that even small dynamical changes in the Rabi splitting can strongly modulate the optical response near the degeneracy. To support our interpretation, we apply TCMT, in which the PS is treated as a system supporting two excitonic resonances that interact with a dispersive photonic mode. To capture the behavior of the system following a pump pulse, we model the resulting slight bleaching with a coupling rate reduced by 0.1% between the waveguide and the excitonic transitions (see SI **Figure SI-3**). While simple, this steady-state approach allows us to capture qualitatively the change in reflectance observed in the distinct transient responses of the PS that we observe experimentally, for each of the different detuning energies from the exciton resonance. Specifically, the simulated spectra reproduce the independent response of the photonic and excitonic components for red and blue energy detuning (see **Figure 3b** and **3f**), but also the reflectivity modulation near the degeneracy point (see **Figure 3d**). The shape of the reflectance modulation near the degeneracy ($\delta \approx 0$) indicates that the WG and energetically broad excitonic modes act as strongly coupled resonances with Rabi splitting, which bleaching reduces due to the pump.

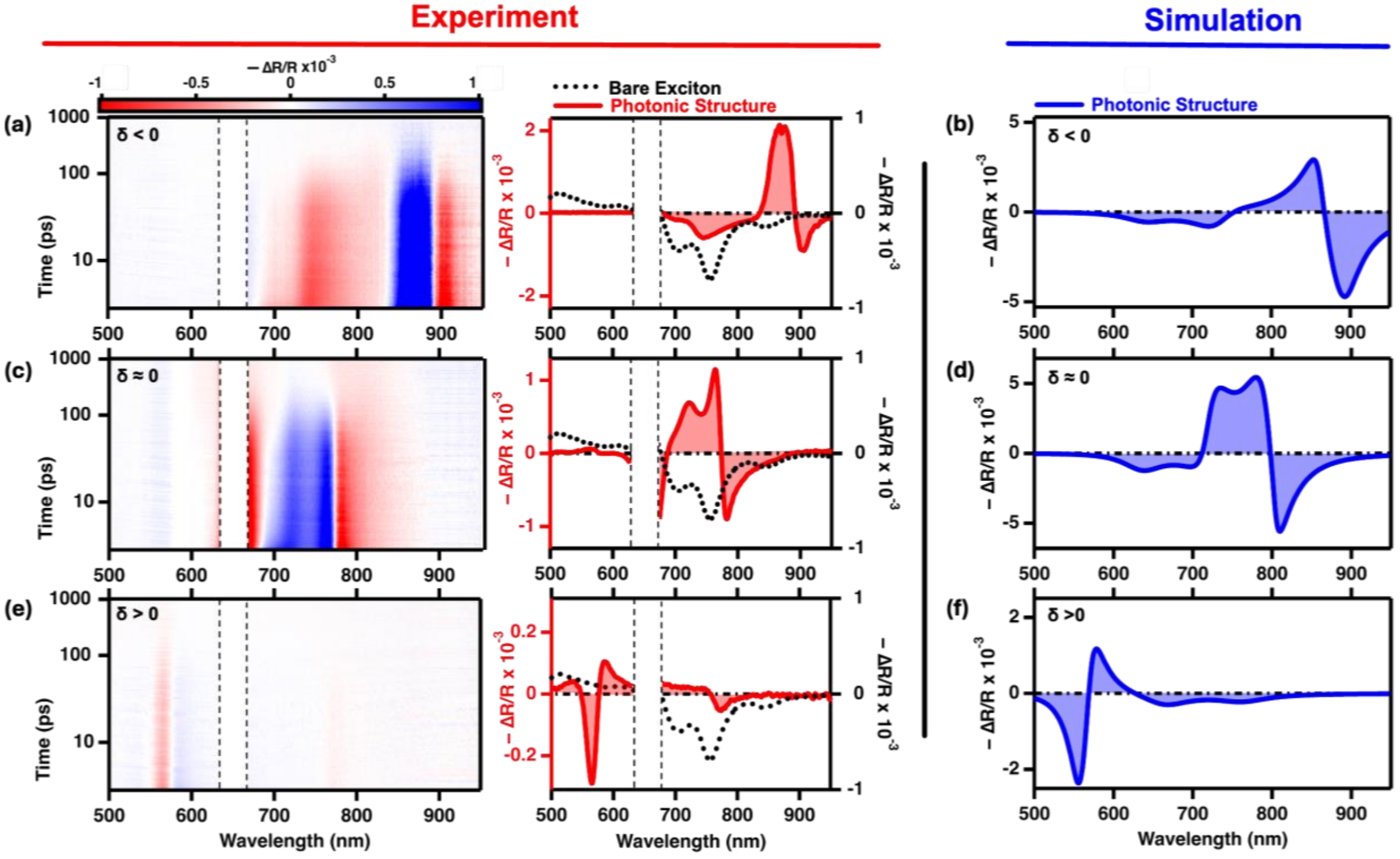

**Figure 3:** Experimental and calculated transient optical data of the PS under excitation resonant with the bare exciton transition at 650 nm (Dotted box – data excluded data due to pump scatter). The data are obtained for **(a)** red ($\delta > 0$), **(c)** resonant ($\delta \approx 0$), and **(e)** blue ($\delta < 0$) detuning energies, demonstrating that the transient respond of the system varies significantly with the detuning energy of the WG mode and the exciton resonance. Next, transient reflection spectra (red color) extracted at the corresponding key delay times, which are depicted along with the bare exciton experimental transient spectra (black color). In the right panel, the simulated transient reflectance using TCMT for the three different energy detuning regions **(b)** $\delta > 0$, **(d)** $\delta \approx 0$, and **(f)** $\delta < 0$.

In support of our model, we show that energy exchange between the exciton and photonic resonance occurs for excitation of either component. In other words, the dynamics of the system near degeneracy depends only on the probe energy and not on the pump energy. To rule out any dependence of the transient response of our PS on the optical pump pulses, we performed transient reflection measurements altering the energy and the in-plane momentum of pump beam ($E_{pump}$, $k_{||-pump}$). First, we photoexcite the PS with lower-energy photons resonant with the first exciton transition $S_0 \rightarrow S_1$ transition ($\lambda_{pump}$ = 750 nm), while keeping the pump and probe in-plane momenta unchanged ($\Delta$ = 4°), (see **Figure SI-14**). Subsequently, we increase the angular offset between pump and probe beams ($\Delta$ = 16°) and set the pump pulses energy in resonance with the second electronic transition ($\lambda_{pump}$ = 650 nm) (see **Figure SI-15**). The nearly identical results were obtained under various photoexcitation conditions ensuring that the observed transient response of the PS is independent from the pump pulses and originates exclusively from the interaction through coupling between the exciton and the photonic mode. The observation of consistent transient spectra obtained using different pump resonances energies confirms that both electronic transitions of the system interact with the WG mode through the coupling.

We also directly photoexcite the WG mode at $\lambda_{pump}$ = 540 nm, where its absorbance, calculated using the TMM to be 0.92, dominates over the bare dielectric layer. As has been previously shown, selective pumping of the dispersive photonic mode can be achieved by tuning the energy and the in-plane wavevector of the pump pulses to match the dispersion curve of the WG mode (phase-matching condition or PMC).[55, 69] For $\lambda_{pump}$ = 540 nm and an angular offset between pump and probe beams of $\Delta$ = 6°, we directly probe the wavevector $k_{||-probe}$ ($\theta_{probe} \approx 54°$) for which the molecular exciton and WG mode energies are nearly degenerate ($\delta \approx 0$). We verified that detuning the energy or wavevector of the pump pulses, keeping the rest experimental conditions unchanged, results in a much weaker transient response (**Figure SI-16**), indicating that the transient signal originates from selective population of the WG states.[56] The resulting transient spectra are identical to those obtained for excitonic resonant photoexcitation when probed near resonance (**Figure 4**), confirming that the photonic and the exciton component respond to one other through a dynamic energy exchange process. This response reveals the characteristic features of the intermediate coupling regime, where the photonic and excitonic mode remain unhybridized, but induce transient signals in one another through internal coupling.

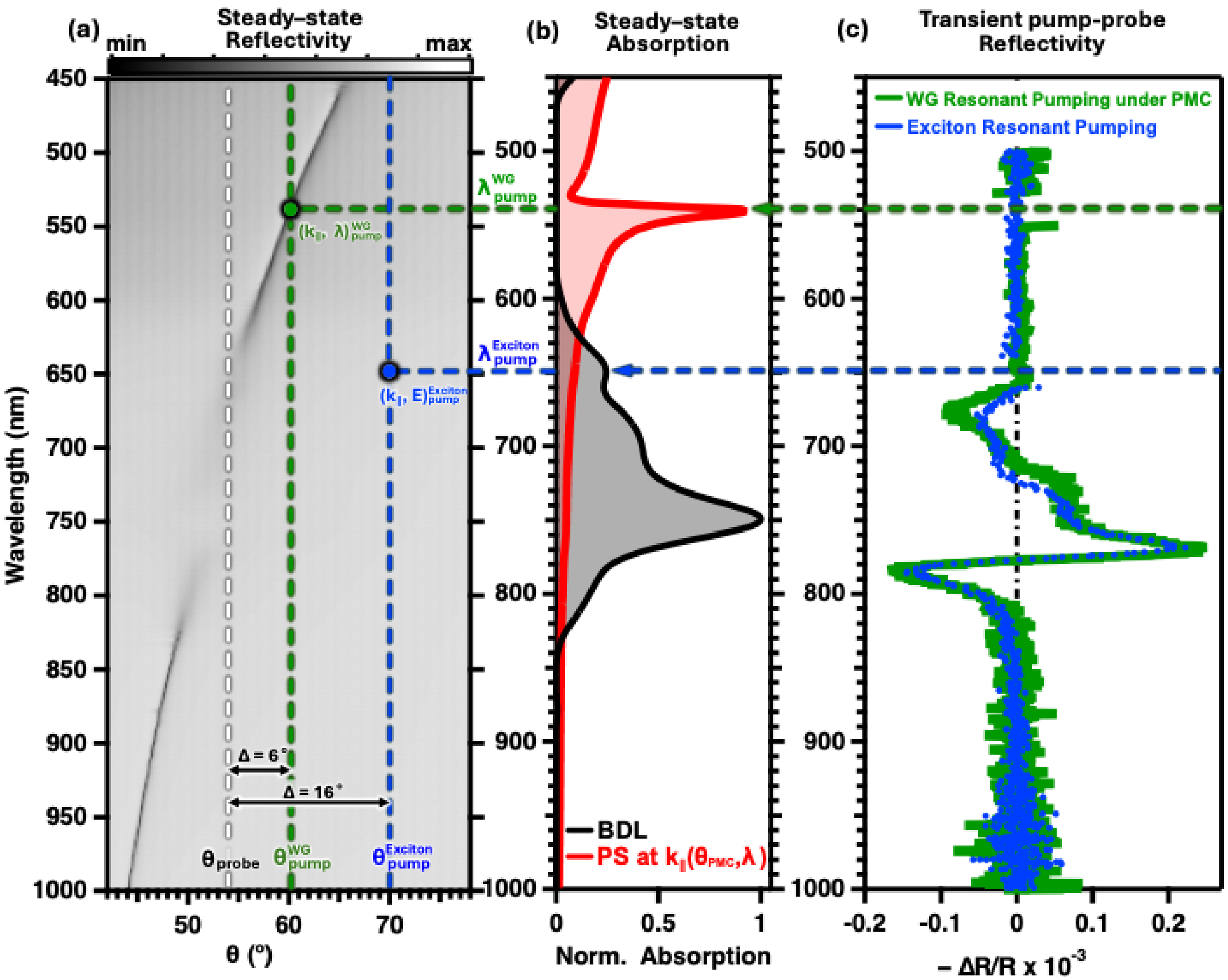

**Figure 4:** Representation of the pump-probe experimental technique, using computational and experimental data, showing the two regimes of photoexcitation. The blue color indicates the data were obtained under the exciton resonant pumping using photons with energy resonant with the molecular exciton electronic transition $S_0 \rightarrow S_2$ ($\lambda_{pump}$ = 650 nm). The green color correspond to measurements were carried out under PMC, in which directly the WG mode is photoexcited with pump pulses matched in energy E ($\lambda_{pump}$ = 540 nm) and the wavevector $k_{||}(\theta_{pump} \approx 60°)$ the WG mode dispersion curve. Both transient measurements were performed under the same probe wavevector $k_{||}(\theta_{probe} \approx 54°)$ which corresponds to WG resonance wavelength $\lambda_{probe} \approx 770$ nm

(δ ≈ 0). This was realized by employing an angular offset between the pump and probe beam of Δ = 16° and Δ = 6° for the excitonic and photonic mode photoexcitation, respectively. **(a)** The calculated angle-reflectivity spectral map, were obtained by modeling the PS using the TMM, provides a representation of the energy-momentum space, highlighting the conditions where the transient reflection experiments were conducted. **(b)** The absorption spectra of the PS at the corresponding $k_{||\text{-}pump}$ for direct WG pumping under PMC using the TMM. The absorption spectra, α (λ), of the bare dielectric layer (black color) extracted using the ellipsometry experimental data (α = 4πk/λ). The absorption spectra confirm that, for each photoexcitation condition, only the corresponding excitonic or photonic component is excited. **(c)** The transient reflection spectra on the PS under excitonic and photonic resonant pumping with pump fluence of 10 u*J*/cm$^2$ and 15 u*J*/cm$^2$ respectively.

**CONCLUSIONS**

Using momentum-resolved transient spectroscopy, we characterize the dynamics of excited states in the intermediate coupling regime. While matter and photonic states remain distinct due to the absence of strong coupling, clear and rapid energy exchange is found to occur near resonance. The effect of this energy exchange is apparent in the transient optical spectra, which exhibit line shapes that are indicative of reduced strong coupling. This behavior is captured using a simple temporal coupled oscillator model with three states. This regime also offers significant opportunities for optical signal modulation. By choosing different momentum vectors, we observe either net loss (away from resonance) or gain (near resonance) at wavelengths corresponding to the bare exciton transitions. We also note that much of the practical benefit from organic strong coupling has enabled by the increased delocalization and transport lengths of the hybrid particles, properties which are inherited from their photonic component. This work suggests that strong coupling may not be fully required to achieve this benefit. It is possible that long-range energy and charge transfer may also be facilitated by intermediate coupling; studies of these phenomena are ongoing. The consequence of this for organic materials should not be minimized as many chromophores feature broad linewidths and/or lower oscillator strengths that preclude an unambiguous demonstration of strong coupling.

**Corresponding Author** - * msfeir@gatech.edu

**Author Contributions**

**EM:** Conceptualization, Methodology, Formal Analysis, Investigation, Writing - Original draft preparation, Writing - Reviewing and Editing, Visualization. **SAM:** Methodology, Formal Analysis, Writing - Original draft preparation, Writing - Reviewing and Editing, Visualization. **KR:** Methodology, Investigation, Writing - Reviewing and Editing. **CL:** Resources, Writing - Reviewing and Editing. **VMM:** Resources, Writing - Reviewing and Editing, Supervision. **AA:** Methodology, Formal Analysis, Writing, Supervision. **MYS:** Conceptualization, Methodology, Formal Analysis, Writing - Original draft preparation, Writing, Supervision. All authors have given approval to the final version of the manuscript.

**Funding**

This work was supported by the Gordon and Betty Moore Foundation, grant DOI 10.37807/GBMF12235. Aspects of this work were partially supported by the Designing Materials to Revolutionize and Engineer our Future (DMREF) program under grant agreement DMR-2323666. SAM and AA were supported by the Air Force Office of Scientific Research and the Simons Foundation. VM was supported by the US Air Force Office of Scientific Research-MURI Grant FA9550-22-1-0317.

**Notes**

The authors declare no competing financial interest.

**Acknowledgment**

The authors acknowledge Dr. Maximilian H. Schreck for the synthesis of the squaraine molecule pySQBcis presented in this work.

**Abbreviations**

Near-Infrared – NIR, Temporal Coupled-Mode Theory – TCMT, Transfer Matrix Method – TMM, Photonic System – PS, Waveguide – WG, Phase Matching Conditions – PMC

TABLE OF CONTENTS FIGURE

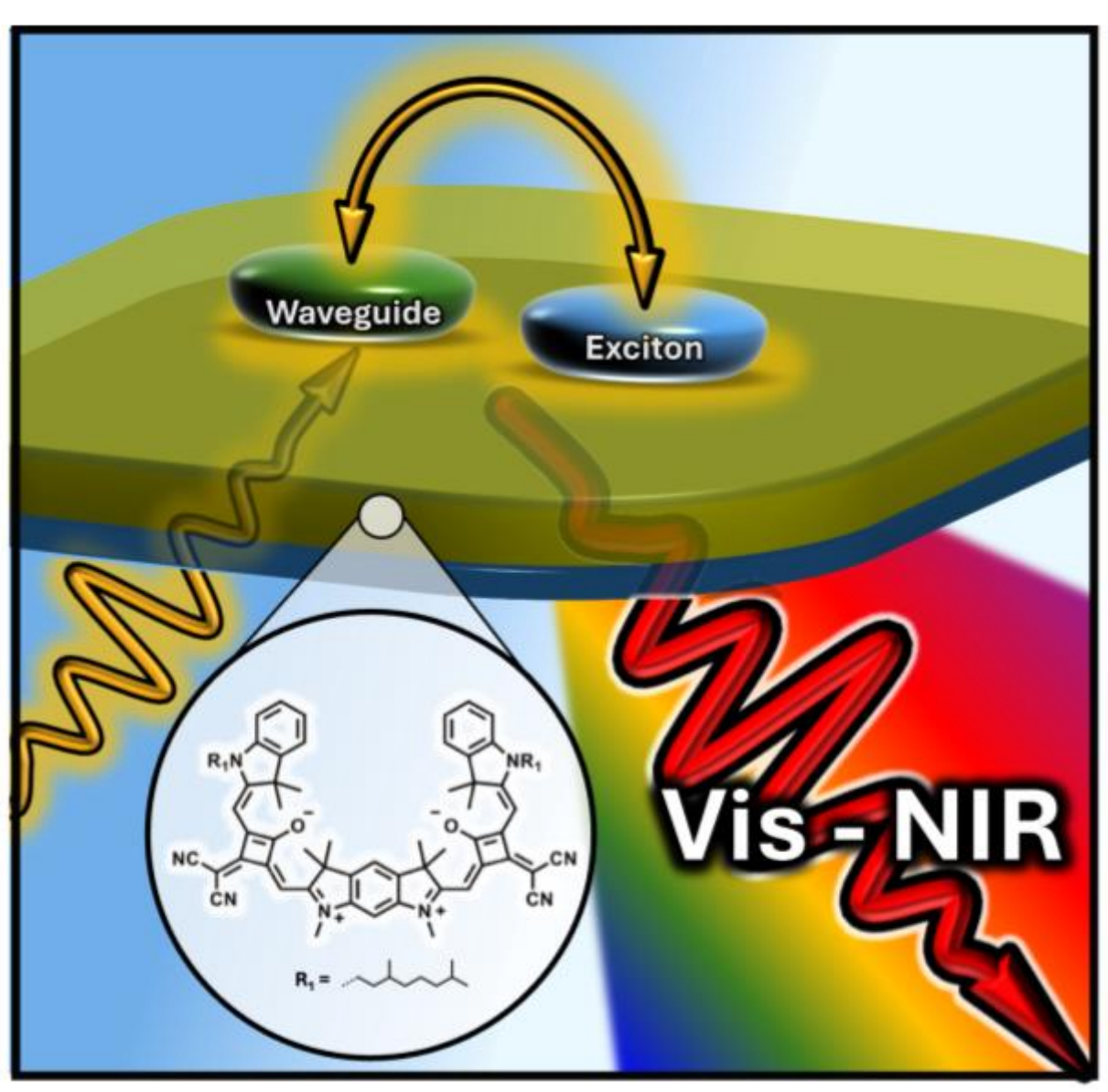